\documentclass[aps,prb,showpacs,twocolumn,superscriptaddress]{revtex4}

\usepackage[utf8]{inputenc}
\usepackage{amsmath}
\usepackage{amsfonts}
\usepackage{graphicx}

\begin{document}

\title{Role of anisotropy in the F\"orster energy transfer from a semiconductor quantum well to an organic crystalline overlayer}

\author{S. Kawka}

\affiliation{Scuola Normale Superiore, Piazza dei Cavalieri, 56126 Pisa, Italy}

\author{G.C. La Rocca}

\affiliation{Scuola Normale Superiore and CNISM, Piazza dei Cavalieri, 56126 Pisa, Italy}


\begin{abstract}
We consider the non-radiative resonant energy transfer from a two-dimensional Wannier exciton (donor) to a Frenkel exciton of a molecular crystal overlayer (acceptor).  We characterize the effect of the optical anisotropy of the organic subsystem on this process. Using realistic values of material parameters, we show that it is possible to change the transfer rate within typically a factor of two depending on the orientation of the crystalline overlayer. The resonant matching of donor and acceptor energies is also partly tunable via the organic crystal orientation.
\end{abstract}


\maketitle

\section{Introduction}

A large effort has been devoted to the successful development of organic light emitting diodes (OLED) despite their intrinsic limitations, particularly regarding carrier injection and transport, as compared to inorganic semiconductor devices. Inorganic-organic hybrid materials may allow the rules that apply to matter and light to be further stretched, potentially seeding a new paradigm in optoelectronic devices taking advantage of the best of both worlds\cite{oulton}. This is of paramount relevance in the strong coupling regime in which novel hybrid quasiparticles are formed, but also, in the weak coupling regime in which Wannier and Frenkel excitons maintain their individuality, an hybrid system offers significant advantages\cite{myrev}. In particular, a way to circumvent the drawbacks of organic materials mentioned above is to use an heterostructure containing an inorganic semiconductor subsystem in which carriers are electrically injected, transported and bound into excitons, coupled to an organic light emitting subsystem via a F\"orster energy transfer process\cite{agranled}. As a step in this direction, it has been proposed and recently demonstrated that  non-radiative energy transfer can be efficient enough (see  Ref.\onlinecite{agraCR} for a review of relevant work).
Following early theoretical predictions\cite{agra97,agra99,denis,denisdot},   the energy transfer process has been observed  from  a quantum well  to a quantum dot overlayer\cite{klimov,zhang2007}, and from a quantum well to an organic overlayer, see \onlinecite{heliotis2006,blumstengel2006,chanyawadee2008} to name a few.

We deal here with such hybrid systems structured in a planar geometry whereby a Wannier exciton in an inorganic semiconductor quantum well plays the role of the donor and a Frenkel exciton in a crystalline organic overlayer that of the acceptor. While many organic subsystems used so far are not crystalline and effectively isotropic, 
most of the organic materials of interest are strongly anisotropic if grown as oriented single crystals. Previous theoretical work has only considered the isotropic case and here we extend it to include the optical anisotropy of the organic subsystem. This will provide a complete quantitative description for the case in which the organic acceptor layer is an oriented single crystal and will allow to characterize how the energy transfer process depends on the orientation of its principal axes.
In section \ref{model}, we describe the theory of  F\"orster energy transfer in a planar hybrid nanostructure taking into due account  optical anisotropy. In section \ref{diel}, we employ a model optical dielectric tensor and parameter values of typical organic media to calculate the transfer rate for various configurations, and we discuss the role of anisotropy.

\section{Theoretical model}
\label{model}

\begin{figure}[!h]
\centering
\includegraphics[width=70mm]{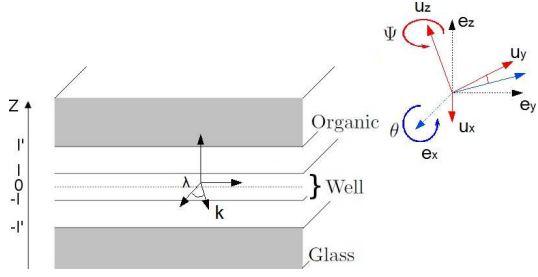}
\caption{Sketch of the planar hybrid heterostructure. The inset shows the angles used to define the orientation of the organic crystal in the top layer (see text for details). (Color online)}
\label{schema}
\end{figure}

We consider here the planar architecture shown in Fig.\ref{schema}. The donor subsystem consists of a semiconductor quantum well of thickness $2l$ sandwiched between two semiconductor barriers of thickness $(l'-l)$. For simplicity, we take the same background dielectric constant $\varepsilon_b$ for the well and the barriers and  we  assume the barriers to be infinitely high (i.e., the Wannier exciton is fully contained in the well region). At the bottom ($z<-l'$) lies a transparent glass substrate with dielectric constant $\varepsilon_g$ while at the top  ($z>l'$) lies the acceptor subsystem consisting of a crystalline organic medium with dielectric constant $\tilde{\varepsilon}_{ij}$. Both of them are supposed to be semi-infinite. The quantities $\varepsilon_b$ and $\varepsilon_g$ include only the contribution of higher resonances (with respect to the exciton energies under consideration) and we consider them to be real. The quantity $\tilde{\varepsilon}_{ij}$ is the total dielectric function of the anisotropic organic material, including in particular the resonant absorption due to the Frenkel excitons, and thus it is a complex valued tensor.

Here we use the same theoretical framework to calculate the F\"orster energy transfer rate in a hybrid nanostructure as in references \onlinecite{agra97,myrev}, and discussed in detail in reference \onlinecite{denis}. This approach is equivalent to the usual F\"orster theory based on the dipole-dipole interaction\cite{agra} and leads to a macroscopic, semiclassical description of the energy transfer. In short, the transfer rate is obtained from the Joule losses suffered  in the organic medium by the electric field induced by the quantum well exciton. As a matter of fact, the presence of the Wannier exciton of energy $E_{exc}=\hbar\omega$ gives rise to a source term in the quantum well corresponding to 
the exciton polarization oscillating at frequency $\omega$ which can be written as:
\begin{align}
\mathbf{P}(\mathbf{r})\,e^{-i\omega t}=\mathbf{d}^{vc} \psi(\mathbf{r},\mathbf{r})\,e^{-i\omega t},
\end{align}  
where $\mathbf{d}^{vc}$ is the matrix element of the electric dipole moment between the Bloch functions of the conduction and valence band extrema and $\psi(\mathbf{r}_e,\mathbf{r}_h)$ is the envelope function describing the bound electron-hole pair ($\mathbf{r}_e$, $\mathbf{r}_h$ being the electron and hole coordinates). 
In our simple model of a 2D Wannier-Mott exciton, the polarization is given by the product of the $1s$-wave function of the 2D relative motion of the electron-hole pair with the lowest subband envelope functions of electron and hole (which are equal in the approximation of infinitely deep well) and with the plane-wave function of the center-of-mass motion. Taking into account their corresponding normalization we can write:
\begin{align}\label{polarization}
\mathbf{P}(\mathbf{r})=\mathbf{d}^{vc} \sqrt{\frac{2}{\pi a_B^2}}\frac{1}{l}\cos^2\left(\frac{\pi z}{2l}\right)\frac{e^{i\mathbf{K}\mathbf{r}_\parallel}}{\sqrt{S}},
\end{align}  
where $S$ is the in-plane normalization area, $\mathbf{K}$ the in-plane wave vector of the center-of-mass motion, $\mathbf{r}_\parallel=(x,y)$ the in-plane component of $\mathbf{r}$ and $a_B$ is the 2D exciton Bohr radius.

Then we look for the electric field $\mathbf{E}(\mathbf{r})\,e^{-i\omega t}$ resulting from this polarization. Having in mind a distribution of Wannier excitons having a broad range of wavevectors as obtained by non resonant pumping or electrical injection, or simply a thermal distribution corresponding to a temperature of the order of 100 K,  we can neglect retardation effects in Maxwell's equations ($k>>\omega/c$ limit) and  use an electrostatic approximation\cite{denis}. The electric field can be written as $\mathbf{E}(\mathbf{r})=-\mathbf{\nabla}\varphi(\mathbf{r})$ and is found from the Poisson equation for the potential $\varphi(\mathbf{r})$:
\begin{equation}\label{poissontot}
\mathbf{\nabla}_i  \varepsilon_{ij}(z) \mathbf{\nabla}_j \varphi(\mathbf{r}) = - 4\pi \rho(\mathbf{r})
\end{equation}
where $\rho(\mathbf{r})=-\mathbf{\nabla}\cdot \mathbf{P}(\mathbf{r})$ is the source charge density, and all the monochromatic oscillating factors $e^{-i\omega t}$ will not be explicitly indicated. The dielectric tensor $\varepsilon_{ij}(z)$ is piecewise constant corresponding to each different layer. We choose it to be isotropic (and real) in the inorganic part: $\varepsilon_{ij}(z)= \varepsilon_g \delta_{ij}$ at $z<-l'$, $\varepsilon_{ij}(z)= \varepsilon_b \delta_{ij}$ at $-l'<z<l'$ and anisotropic (and complex) in the organic layer: at $z>l'$, $\varepsilon_{ij}(z)= \tilde{\varepsilon}_{ij}$. 
The appropriate boundary conditions at the interfaces are the continuity of the tangential electric field and of the normal electric displacement between each layer.
Knowing the electric field we can calculate the transfer rate (inverse transfer time) of the excitation into the organic medium:
 \begin{align}\label{taugene}
\frac{1}{\tau}=\frac{1}{2\pi\hbar} \int_{z>l'} \mathrm{Im}(\tilde{\varepsilon}_{ij}) \left(E_{i}(\mathbf{r})\right)^* E_j(\mathbf{r}) d^3r.
\end{align}  
This expression is equivalent to applying the Fermi Golden Rule to the decay of one excited state in the quantum well into the excited states of the organic molecules in the linear regime approximation (for a derivation of Eq.\ref{taugene} in terms of a microscopic inelastic scattering rate of Wannier excitons due to resonant two-level molecules in the organic medium see Ref.\onlinecite{denisigma}). 
The power $W$ dissipated in the organic layer is given by $W=\hbar\omega/\tau$.
Such energy transfer mechanism have been shown to be fast enough to efficiently quench the Wannier exciton luminescence and to turn on the organic molecule light emission. While previous theoretical work has only considered the case of an effectively isotropic organic medium (i.e.,  $\tilde{\varepsilon}_{ij} \propto \delta_{ij}$), we  focus here on the effects of the anisotropy of the organic layer on the transfer time $\tau$, assuming for the donor subsystem the same model appropriate to zincblende semiconductors used there\cite{denis}.

Due to the in-plane translational symmetry of the source (we are dealing with free excitons in the well), we consider the polarization for a given in-plane wave vector. In this case three modes of different symmetry can be identified: longitudinal (L) where $\mathbf{d}^{vc}$ is along the in-plane wave vector, perpendicular (Z) where the dipole moment is oriented along the z-axis, and transverse (T) for which the polarization is orthogonal to the two first case. We notice that the T mode does not give rise to any charge density and thus the dipole-dipole interaction vanishes. The two remaining polarizations lead respectively to two different charge densities in the quantum well $\rho(\mathbf{r}) = \rho^{(L,Z)}(z)e^{i\mathbf{K}\mathbf{r}_\parallel} $ , given by:
\begin{align}
\rho^{(L)} (z)&=-iKl\rho_0(1+\cos qz) \\
\rho^{(Z)}(z) &=ql\rho_0\sin qz \\
{\rm with }\quad \rho_0 &=\frac{1}{\sqrt{2\pi a_B^2}}\frac{d^{vc}}{\sqrt{S}l^2} \quad,\quad q=\pi /l \quad .
\end{align}
For any given source exciton symmetry (L or Z, which will not be explicitly indicated), writing $\varphi(\mathbf{r}) = \phi(z)e^{i\mathbf{K}\mathbf{r}_\parallel} $ the Poisson equation gives
\begin{equation}\label{poisson}
 \left\{
          \begin{array}{lr}
           \left(\frac{d^2}{dz^2} - K^2\right) \phi(z) = - 4\pi \rho(z)/\varepsilon_b & |z|< l \\
           \left(\frac{d^2}{dz^2} - K^2\right) \phi(z) = 0 &  l<|z|< l'\\
	  \left(\frac{d^2}{dz^2} - K^2\right) \phi(z) = 0 &  z<-l'\\
               \multicolumn{2}{c}{\begin{array}{l}           
\Big(\varepsilon_{zz} \frac{d^2}{dz^2} +i \left[ (\varepsilon_{zx}+\varepsilon_{xz})K_x+(\varepsilon_{zy}+\varepsilon_{yz})K_y \right] \frac{d}{dz}  \\  - \left[ \varepsilon_{xx} K_x^2 + \varepsilon_{yy} K_y^2  + (\varepsilon_{xy} + \varepsilon_{xy}) K_xK_y \right] \Big) \phi(z) = 0 
                \end{array} }  \\
  &    z>l' \\
          \end{array}
        \right.
\end{equation}
In the organic part $\phi$ has the form $\phi(z)=\rho_0\, C \,e^{\gamma K(z-l')}$ with $C$ an amplitude coefficient and $\gamma$ a complex number. The value of $\gamma$ is the solution of the last equation in (\ref{poisson}) such that $Re(\gamma)<0$ to satisfy the boundary condition at infinity: 
\begin{align}
&\nonumber \varepsilon_{zz} \gamma^2  +i \left[ 2\varepsilon_{zx}\frac{K_x}{K}+2\varepsilon_{zy}\frac{K_y}{K} \right] \gamma \\
& - \left[ \varepsilon_{xx} \frac{K_x^2}{K^2} + \varepsilon_{yy} \frac{K_y^2}{K^2}  + 2\varepsilon_{xy}  \frac{K_xK_y}{K^2} \right] =0 \quad;
\end{align}
consequently $\gamma$ depends on $\mathbf{K}$ and on the orientation of the organic material. Solving the 8 by 8 system obtained imposing the two boundary conditions on each of the four interfaces (at $z=\pm l$ and $z=\pm (l+l')$), we find for the two source modes, respectively, the amplitude coefficients:
\begin{align}
\nonumber C_{\mathbf{K}}^{(L)} =& - \frac{i8\pi^2q}{K(K^2+q^2)}\times \\
& \frac{ \sinh(Kl) \left(\varepsilon_b\cosh(Kl') + \varepsilon_g\sinh(Kl') \right) }{(\varepsilon_g +\tilde{\varepsilon})\varepsilon_b\cosh(2Kl') + (\varepsilon_g\tilde{\varepsilon}+\varepsilon_b^2)\sinh(2Kl')} \\
\nonumber  C_{\mathbf{K}}^{(Z)} =&  \frac{8\pi^2q}{K(K^2+q^2)} \times \\
&  \frac{ \sinh(Kl) \left(\varepsilon_g\cosh(Kl') + \varepsilon_b\sinh(Kl') \right) }{(\varepsilon_g +\tilde{\varepsilon})\varepsilon_b\cosh(2Kl') + (\varepsilon_g\tilde{\varepsilon}+\varepsilon_b^2)\sinh(2Kl')}
\end{align}
where the parameter $\tilde{\varepsilon}=-\tilde{\varepsilon}_{zz} \gamma -i \left( \varepsilon_{zx}\frac{K_x}{K}+\varepsilon_{zy}\frac{K_y}{K} \right)$ comes from the continuity of the normal electric displacement at the boundary between the organic and inorganic layers. These amplitude coefficients depend on $\mathbf{K}$ and on the orientation of the organic material in a rather involved way through $\tilde{\varepsilon}$ and $\gamma$. We recover previous results for the isotropic case \cite{denis} provided that $\tilde{\epsilon}_{ij} \propto \delta_{ij} $ and $\epsilon_g = \tilde{\epsilon}$.

The electric field in the organic layer is then:
\begin{equation}
\mathbf{E}(\mathbf{r}) = [ -i\mathbf{K} - \gamma K \mathbf{e}_z ] \varphi(\mathbf{r})
\end{equation}
where $\mathbf{e}_z$ is the unit vector in the $z$ direction. Finally, from Eq. (\ref{taugene}),  the transfer time $\tau$ for an exciton of in-plane wave vector $\mathbf{K}=K\,(\cos\lambda,\sin\lambda)$ is given by
\begin{align}\label{lambdatau}
 \nonumber  \frac{1}{\tau(K,\lambda)} =& \frac{|d^{vc}|^2}{2\pi^2\hbar a_B^2}\frac{K|C_K|^2}{(2l)^4} \frac{-1}{\mathrm{Re}\gamma} \times\\
& \nonumber \bigg[ \mathrm{Im}\varepsilon_{xx} \cos^2\lambda + \mathrm{Im}\varepsilon_{yy} \sin^2\lambda  +\mathrm{Im}\varepsilon_{zz}|\gamma|^2 \\ 
& \nonumber +2\mathrm{Im}\gamma(\mathrm{Im}\varepsilon_{xz}\cos\lambda +\mathrm{Im}\varepsilon_{yz}\sin\lambda) \\
&+2\mathrm{Im}\varepsilon_{xy}\cos\lambda\sin\lambda \bigg] 
\end{align}

The increased complexity of the present analytical results with respect to previous theoretical work\cite{denis} is due partly to having lifted the mirror symmetry with respect to the $z=0$ plane assumed there,
and partly to having included an anisotropic dielectric tensor for the acceptor subsystem. The latter point gives rise to significant consequences on the energy transfer process as discussed below, while the former point allows to describe the quantitative dependence of the transfer time on the value of the substrate dielectric constant which, however, within a reasonable range of parameter values is a minor effect.
In most cases of experimental interest, all directions for the center of mass wave vector of the exciton in the well are equiprobable (i.e., the exciton distribution has cylindrical symmetry), and an average over the angle $\lambda$ is taken, which is done numerically. For a thermalized population of quantum well excitons,  we also average over the energy  $\frac{\hbar^2 K^2}{2M}$, with $M$ the exciton mass,  according to the Boltzmann distribution. Taking these averages tends to reduce the effect of the anisotropy, and this would also be the case for the energy transfer of a localized exciton, in which case the wave vector distribution is given by the Fourier transform of the localised wave function of the center of mass motion.

\section{Results and discussion}\label{diel}

For illustrative purposes, we will focus on the effects of the anisotropy and thus only change the organic overlayer configuration keeping all the rest of the heterostructure fixed, and similar to that assumed in previous theoretical work. We will thus consider a variety of cases of anisotropic organic materials without bothering to select in each case an appropriately matched donor subsystem. In this way, the effects of the anisotropy will be singled out and discussed. For the donor subsystem, we take a II-VI semiconductor quantum well (e.g. ZnSe) for which typically\cite{cingolani} $\varepsilon_b \simeq 6$, $\mathbf{d}^{vc} \simeq 12 ea_B$ and $a_B$ is taken to be $25 \mathring{A}$, $M=0.76m_0$. For the structure,we take $l=30 \mathring{A}$ and $l'=40 \mathring{A}$, while the glass substrate and the organic top layer are infinite in the z-direction.

The dielectric function of an anisotropic organic material can be written as \cite{agra84}:
\begin{equation}
\varepsilon_{ij}=\varepsilon_{\infty}\delta_{ij} + \frac{8\pi}{V} \sum_n \frac{d_{n,i}d_{n,j} \hbar\omega_n}{(\hbar\omega_n)^2-(\hbar\omega)^2-i\Gamma_n\hbar\omega}
\end{equation}
where $\varepsilon_{\infty}$ is the isotropic high-frequency dielectric constant, $V$ is the volume of the lattice cell, $d_{n,i}$ is the $i$th component of the transition dipole moment of the $n$th excitonic eigenstate with energy $\hbar\omega_n$ and damping $\Gamma_n$.  In simple cases, it is possible to use an effective model \cite{tavazzi08}, typically near the exciton resonances, where the optical response is modeled by a real background constant and several Lorentz transitions for each principal axis $j$, described by their energy $E_{j,0}$, coupling amplitude $A_j$ and damping $\Gamma_j$:
\begin{equation}\label{ej}
\varepsilon_{j}(\omega)=\varepsilon_{j,\infty} + \frac{A_j\Gamma_j E_{j,0}}{E_{j,0}^2-(\hbar\omega)^2 - i\Gamma_j \hbar\omega}
\end{equation}
$\varepsilon_{j,\infty}$ is here the background constant adapted to each axis. The values of the parameters are obtained by fitting experimental data. Here we use such a simple model with experimental values for the energy, coupling amplitude and damping of each relevant exciton transition, see Tables \ref{tabletetra},\ref{tablebuta}.

\begin{table}[ht]
\caption{Experimental values obtained for tetracene from the fitting of ellipsometry data \cite{tavazzi08}, valid for an energy range from 2.2eV to 2.5eV. }
\centering 
\begin{tabular*}{0.45\textwidth}{@{\extracolsep{\fill}} l l l l} 
\hline\hline 
 & $\varepsilon_1$ & $\varepsilon_2$ & $\varepsilon_3$ \\ [0.5ex] 
\hline 
  $\varepsilon_\infty$ & $1.39$ & $1.00$ & $2.10$ \\
 $E_0$ & 2.38 eV & 2.46 eV & \\ 
  A & 2.971 & 0.391 & \\ 
  $\Gamma$ & 0.088 eV & 0.057 eV & \\
 [1ex] 
\hline 
\end{tabular*}
\label{tabletetra}
\end{table}

\begin{table}[ht]
\caption{Experimental values obtained for $\alpha-$tetraphenyl-butadiene (TPB) from the fitting of ellipsometry data \cite{tavazzi11} (converting the Gaussian fit into the Lorentz model, keeping the same peak value and area under the peak of the exciton resonance), valid for an energy range from 3.2eV to 4.2eV.}
\centering 
\begin{tabular*}{0.45\textwidth}{@{\extracolsep{\fill}} l l l l} 
\hline\hline 
 & $\varepsilon_1$ & $\varepsilon_2$ & $\varepsilon_3$ \\ [0.5ex] 
\hline
  $\varepsilon_\infty$ & 3.1 & 2.84 & $2.37$ \\
 $E_0$ & 3.83 eV & 3.61 eV & \\ 
  A & 2.46 & 1.73 & \\ 
  $\Gamma$ & 0.59 eV & 0.58 eV & \\ 
 [1ex] 
\hline 
\end{tabular*}
\label{tablebuta}
\end{table}

The orientation of the organic crystal is given by three Euler's angles $(\Phi,\theta,\Psi)$, as defined in  Ref.~\onlinecite{landau} : a rotation of angle $\Phi$ around the z-axis followed by a rotation of angle $\theta$ around the new x-axis and a rotation of angle $\Psi$ around the last new obtained z-axis. However, due to the rotational invariance of the inorganic subsystem around the z-axis, only the angles $(\theta,\Psi)$ are needed. Besides, due to the mirror symmetry perpendicularly to each axis, it is enough to take them between 0 and $\pi/2$.
The coordinates of the optical axes $\{ \mathbf{u}_i,i=x,y,z\}$ expressed in the device frame $\{ \mathbf{e}_i\}$ (see Fig.\ref{schema}) are given by $\mathbf{u}_i = \mathcal{R} \mathbf{e}_i$ where $\mathcal{R}$ is a rotation matrix constituted by a rotation of angle $\theta$ around $\mathbf{e}_x$ followed by a rotation of angle $\Psi$ around $\mathbf{u}_z$:
\begin{equation}
\nonumber
\mathcal{R} =  \mathcal{R}^{u_z}(\Psi) \mathcal{R}^x(\theta) 
\end{equation}
This rotation can be rewritten in the device frame as $\mathcal{R} =  \mathcal{R}^x(\theta) \mathcal{R}^{z}(\Psi) $. Then the dielectric tensor $\tilde{\varepsilon}$ of the organic compound reads:
\begin{equation}\label{anglestet}
\tilde{\varepsilon}_{ij} = \mathcal{R}^x_{ik}(\theta)  \mathcal{R}^z_{kl}(\Psi) \varepsilon_{l} \mathcal{R}^z_{lm}(-\Psi)  \mathcal{R}^x_{mj}(-\theta)
\end{equation}
where $\varepsilon_{l}$ is given by (\ref{ej}). This orientation is chosen (or fixed) by the growth conditions of the organic component. 
In the following, we discuss the dependence of the transfer time on the orientation of the organic crystal, and the optimal choice of frequency for the donor for a given orientation.

\begin{figure}[!h]
\centering
\includegraphics[width=70mm]{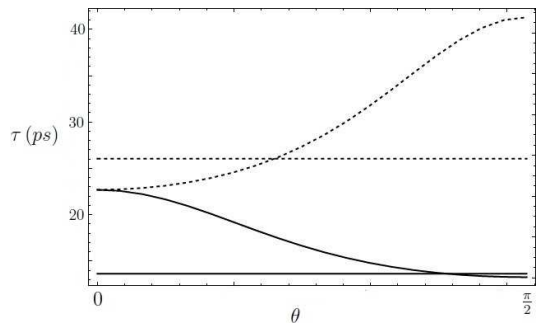}
\caption{Transfer time of a $L$-polarized exciton in function of the orientation for a uniaxial crystal with relative permittivity given by $\varepsilon_{1}=\varepsilon_{2}=2$, $\varepsilon_{3}=2+3i$, dashed curve, and $\varepsilon_{1}=\varepsilon_{2}=2+3i$, $\varepsilon_{3}=2$, solid curve. The straight lines are the mean values associated to the corresponding isotropic case.}
\label{uniaxe}
\end{figure}

We consider a thermal distribution of excitons in the well and  we assume a temperature of 100K. At a given frequency of emission, we calculate the transfer time as a function of the orientation of the crystal. 
At first, we discuss the case of a uniaxial medium. Its orientation is determined by only one angle $\theta$ as defined in (\ref{anglestet}) between the optical axis and the normal of the planar structure. In Fig.\ref{uniaxe} we give the transfer time as a function of $\theta$ for a medium where the optical axis is the absorbing one (dashed curve) or the only non-absorbing one (solid curve). We find a transfer time of the order of a few tens of picoseconds comparable to that of the isotropic case\cite{denis}. This is competitive to any other decay channel for the excitons in the well  proving the efficiency of the energy transfer to the organic media. We observe, in particular, a variation of a factor 1.7 in the transfer time depending on $\theta$, so the efficiency of the transfer is affected by the orientation of the crystal in a significant way. The minimum transfer time is lower in the case of two absorbing axis as expected as dissipation in the organic is increased in this case.

\begin{figure}[!h]
\centering
\includegraphics[width=70mm]{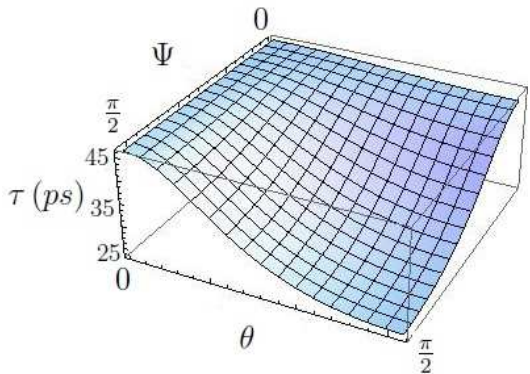}
\caption{Transfer time of a $L$-polarized exciton as a function of the orientation of the tetracene crystal, given by the two angles $\Psi$ and $\theta$, at an energy 2.38 eV. for which $\varepsilon_1=2.59+3.0i, \varepsilon_2=2.21+0.05i ,\varepsilon_3=2.09$. (Color online)}
\label{to3Dtetra}
\end{figure}

\begin{figure}[!h]
\centering
\includegraphics[width=70mm]{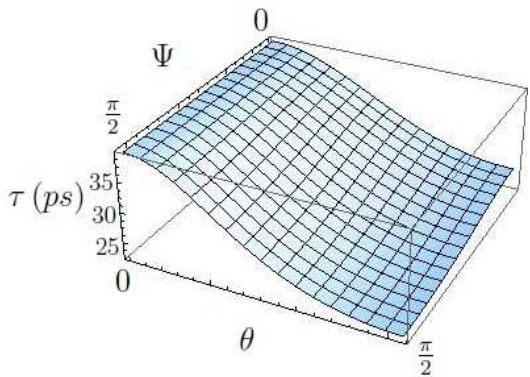}
\caption{Transfer time of a $L$-polarized exciton as a function of the orientation of the $\alpha$-1,4,4-tetraphenyl-$1,3$-butadiene crystal, given by the two angles $\Psi$ and $\theta$, at an energy 3.83 eV for which $\varepsilon_1=3.1+2.46i, \varepsilon_2=2.0+0.95i ,\varepsilon_3=2.37$. (Color online)}
\label{to3Dbu}
\end{figure}

\begin{figure}[!h]
\centering
\includegraphics[width=70mm]{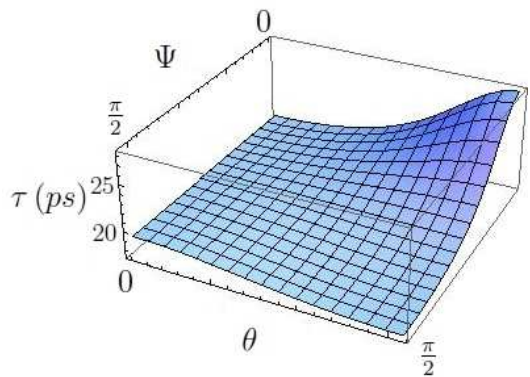}
\caption{Transfer time of a $L$-polarized exciton as a function of the orientation of PTCDA crystal, given by the two angles $\Psi$ and $\theta$, at an energy 2.2 eV. where $\varepsilon_1=2.46+0.15i, \varepsilon_2=7.1+3.41i ,\varepsilon_3=6.3+6.5i$ (ref.~\onlinecite{alonso}). (Color online)}
\label{to3DPTCDA}
\end{figure}

\begin{figure}[!h]
\centering
\includegraphics[width=70mm]{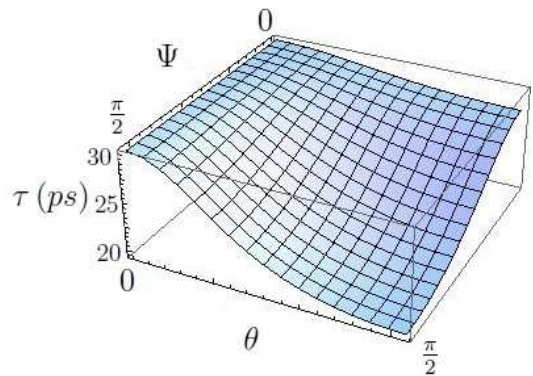}
\caption{Transfer time of a $L$-polarized exciton as a function of the orientation of anthracene crystal, given by the two angles $\Psi$ and $\theta$, at an energy 3.12 eV where $\varepsilon_1=4.71+4.86i ,\varepsilon_2=4.65+0.71i, \varepsilon_3=3.01+0.46i$ (ref.~\onlinecite{alonso}). (Color online)}
\label{to3Dant}
\end{figure}

To illustrate the case of biaxial media, we consider tetracene and $\alpha$-1,4,4-tetraphenyl-$1,3$-butadiene, for which dielectric function data were given above, and anthracene and PTCDA. For both the latter, we can use data obtained directly from ellipsometry mesurement at the relevant exciton energy in Ref.~\onlinecite{alonso}. We show in Fig.~\ref{to3Dtetra},~\ref{to3Dbu},~\ref{to3DPTCDA} and~\ref{to3Dant} their respective transfer times as a function of the orientation of the crystal, defined by the two angles $\Psi$ and $\theta$ (\ref{anglestet}).  The order ofmagnitude of the exciton lifetime is of a few tens of picoseconds, with a minimum around 20~ps. We observe a difference in the transfer time depending on the orientation of a factor~1.8 for tetracene, 1.6 for TPB and PTCDA, and a factor~1.5 for anthracene.
The qualitative shape of these plots depends on the number of dissipative axis. In the case of $\alpha$-1,4,4-tetraphenyl-$1,3$-butadiene, the transfer time is higher when the non-absorbing axis is along the $z$-direction. For tetracene almost all the dissipation is along a given axis and the transfer time  is maximum when one of the two non dissipative axis is along the $z$-direction.
We show the one exciton life-time for a $L$-mode exciton. The $Z$-mode gives nearly the same results. 

\begin{figure}[!h]
\centering
\includegraphics[width=60mm]{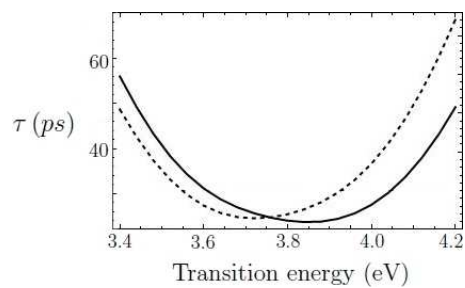}
\caption{Transfer time of a $L$-polarized exciton as a function of the resonant energy for $\alpha$-1,4,4-tetraphenyl-$1,3$-butadiene crystal in two different orientation $\{\theta=\pi/2, \Psi=\pi/2 \}$ thick curve, and $\{\theta=\pi/2,\Psi=0 \}$ dashed curve. The minimums correspond to the two frequencies of absorption (cf Table \ref{tablebuta}) so it is possible to excite predominantly one or the other of the two excitons of the $\alpha$-1,4,4-tetraphenyl-$1,3$-butadiene crystal.}
\label{en}
\end{figure}

These results have to be compared to the case where the organic compound is on a powder form, where the size of a mono-crystal is much smaller than the wavelength of the resonant light. In this case an average over all possible directions for the mono-crystals is realized, which is equivalent of averaging the dielectric tensor. We found a transfer time of 28ps for tetracene, 26ps for TPB, 21ps for anthracene and 17ps for PTCDA. Thus, in general, the transfer efficiency for this effectively isotropic case is comparable to that of a nearly optimally oriented single crystal.

\begin{figure}[!h]
\centering
\includegraphics[width=60mm]{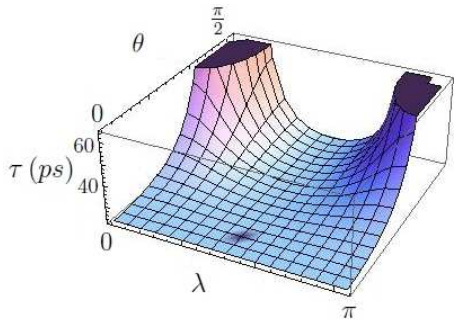}
\caption{Transfer time of a $L$-polarized exciton in function of the direction of this latter and of the orientation $\theta$ of a uniaxial crystal, with relative permittivity given by $\varepsilon_{1}=\varepsilon_{2}=2$, $\varepsilon_{3}=2+3i$. The transfer time is going to infinity for $\{\lambda=0, \pi ; \theta=\pi/2\}$. (Color online)}
\label{lambda}
\end{figure}

For a given orientation of the crystal, as it could be imposed by the growth conditions, one can find a frequency that minimize the exciton life-time. In Fig.\ref{en} we show the transfer time as a function of the resonant energy for two different orientation of the $\alpha$-1,4,4-tetraphenyl-$1,3$-butadiene crystal, each curve presenting a minimum respectively around 3.7eV and 3.9eV which correspond to the two frequencies of absorption of the $\alpha$-1,4,4-tetraphenyl-$1,3$-butadiene, each exciton being excited separately. As a consequence, as a fucntion of the orientation, it possible to improve the transfer efficiency by improving the tuning with the emission frequency of the donor.

Finally, for illustrative purposes, we consider the energy transfer time given by Eq.\ref{lambdatau} for an excitons having a center of mass $\mathbf{K}$ along a preferential direction with a thermalised energy. We observe the variation of the transfer time with that direction given by the angle $\lambda$ and the orientation of the crystal $\theta$. In the case of a uniaxial crystal with absorption in one direction, see Fig.\ref{lambda}, there is no absorption when the Frenkel exciton is perpendicular to the plane defined by the source exciton and the z-axis, then the transfer time is infinite. In the case of tetracene crystal, where there is a second exciton with a smaller oscillator strength, we observe a contrast depending on the orientation of the crystal of a factor of about 40.

\section*{ACKNOWLEDGMENTS}

We acknoledge fruitful discussions with \hbox{V.M. Agranovich}.

This work is supported by the European FP7 ICARUS program, grant agreement no 237900.


\end{document}